\documentclass[aps,prl,twocolumn,showpacs,superscriptaddress,groupedaddress]{revtex4}  
\usepackage{dcolumn}   
\usepackage{bm}        
\usepackage{amssymb}   
\usepackage{acronym}
\usepackage[hidelinks]{hyperref}
\usepackage{amsmath}
\usepackage{subfigure} 

\usepackage{xspace}
\newcommand{\C}{$^{\circ}$C\xspace}

\ifx\pdfoutput\undefined
\usepackage{graphicx}
\else
\usepackage[pdftex]{graphicx}
\usepackage{epstopdf}
\fi

\usepackage[percent]{overpic}

\def\figureWidth{3.30}

\graphicspath{{figures//}}
\usepackage[percent]{overpic}
\usepackage{color}

\begin{document}


\title{All-Optical Switching Demonstration using Two-Photon\\ Absorption and the Classical Zeno Effect}
\author{S. M. Hendrickson}                           
\author{C. N. Weiler}
\affiliation{Applied Physics Laboratory, The Johns Hopkins University, Laurel, Maryland 20723}
\author{R. M. Camacho}
\author{P. T. Rakich}
\author{A. I. Young}
\author{M. J. Shaw}
\affiliation{Sandia National Laboratories, Albuquerque, New Mexico 87185}
\author{T. B. Pittman}
\author{J. D. Franson}
\affiliation{University of Maryland, Baltimore County, Baltimore, Maryland 21250}
\author{B. C. Jacobs}
\affiliation{Applied Physics Laboratory, The Johns Hopkins University, Laurel, Maryland 20723}
\date{\today}

\begin{abstract}

Low-contrast all-optical Zeno switching has been demonstrated in a \acs{SiN} microdisk resonator coupled to a hot atomic vapor.  The device is based on the suppression of the field build-up within a microcavity due to non-degenerate two-photon absorption.  This experiment used one beam in a resonator and one in free-space due to limitations related to device physics.  These results suggest that a similar scheme with both beams resonant in the cavity would correspond to input power levels near 20 nW.  

\end{abstract}


\maketitle


\acrodef{SiN}[Si$_3$N$_4$]{Silicon Nitride}
\acrodef{Rb}{Rubidium}
\acrodef{TPA}{two-photon absorption}
\acrodef{SEM}{scanning electron microscope}
\acrodef{cavRes}[$\lambda_{c}$]{cavity resonance}
\acrodef{780Res}[$\lambda_{Rb}^{780}$]{780 atomic resonance}
\acrodef{1529Res}[$\lambda_{Rb}^{1529}$]{1529 atomic resonance}
\acrodef{HF}{Hydrofluoric acid}
\acrodef{Q}{quality factor}
\acrodef{QZE}{quantum Zeno effect}
\acrodef{IRS}{inverse Raman scattering}



The \ac{QZE} can prevent a randomly occurring process by frequent measurement \cite{misra:756}.  It has previously been shown \cite{PhysRevA.70.062302, Franson:07,PhysRevA.74.053817} that this effect could be used to suppress errors in quantum logic gates using strong \ac{TPA}.  Recently, this work was extended to show that the \ac{QZE} has a classical analog that could be used to create a low-loss all-optical switch \cite{PhysRevA.79.063830} capable of operating at low powers.

Whereas the \ac{QZE} prevents the buildup of a probability amplitude, the classical Zeno effect suppresses the coherent buildup of the electromagnetic field amplitude within a microresonator.  To see how this can be used to create a switch, consider a system in which the resonator is strongly coupled to a two-photon absorbing medium such that two distinct frequencies are required for absorption to take place.  With the resonator critically coupled to two waveguides, the presence of a resonant input at either of the two frequencies will result in the light coupling into the resonator and leaving the opposite waveguide.  This is due to the destructive interference between the light remaining in the waveguide and the built-up field amplitude in the cavity that couples back to the waveguide.  When both frequencies are present in the cavity the \ac{TPA} prevents the coherent intra-cavity field buildup and the input beams pass by the resonator because there is now insufficient amplitude in the cavity to result in interference.

\begin{figure}[h!]
\subfigure[]{\includegraphics[width = \figureWidth in]{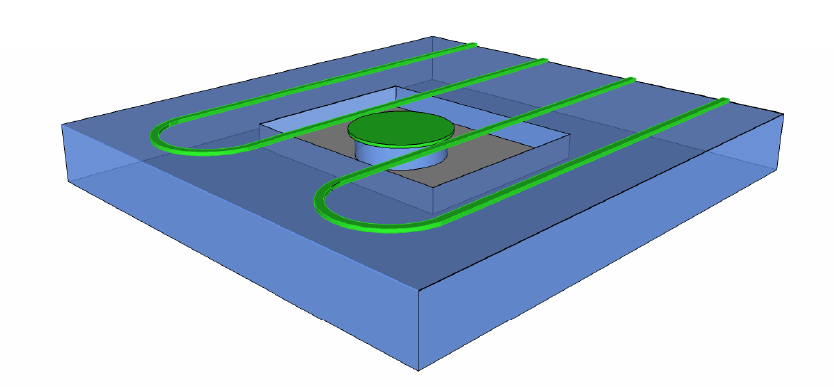}\label{fig:SandiaSchematic}}


\subfigure[]{
\begin{overpic}[width = \figureWidth in]{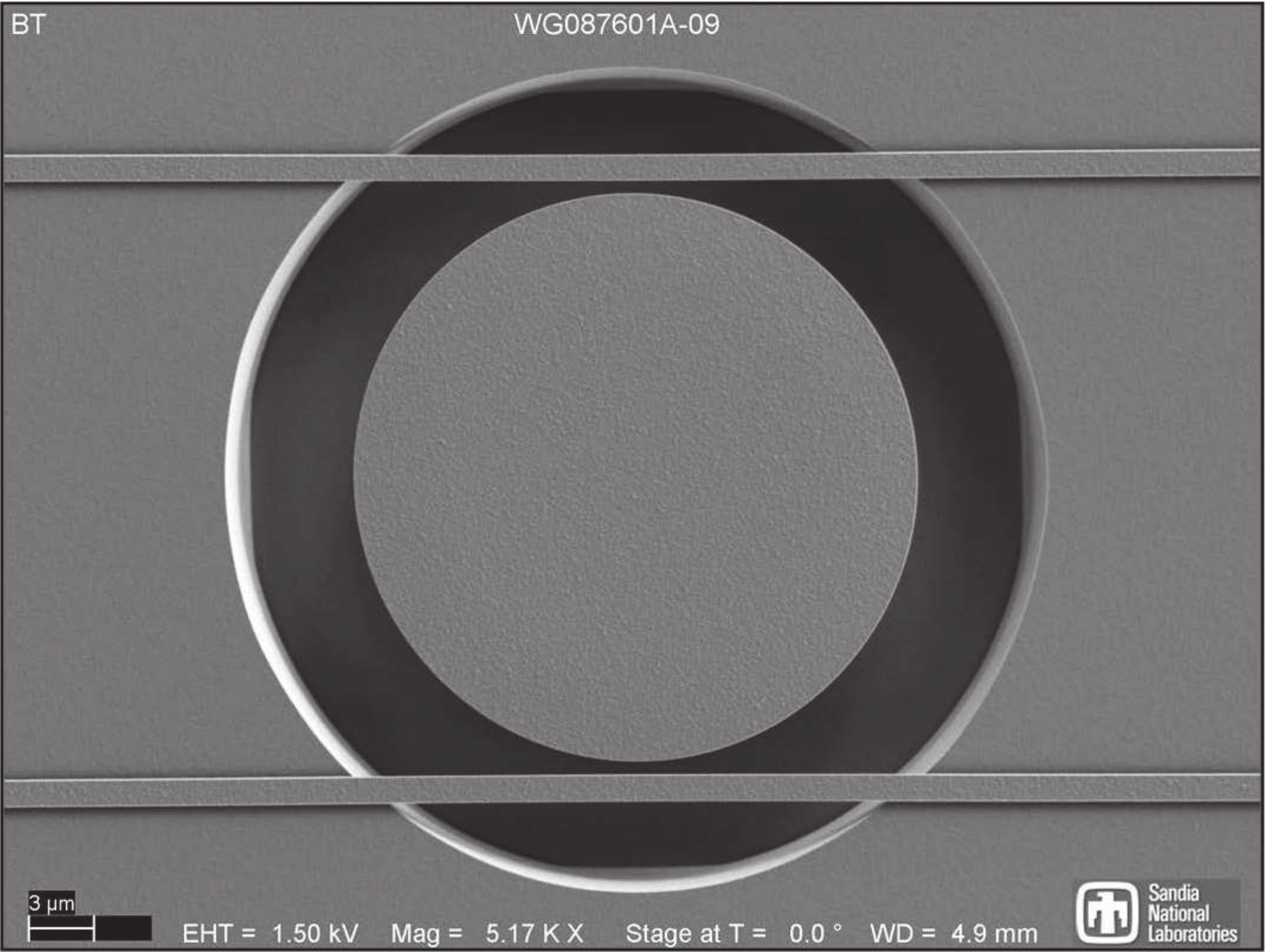}
\put(5,64.5){\textcolor{white}{\Large{P$_{in}$}}}
\put(5,60.8){\textcolor{white}{$\longrightarrow$}}

\put(5,15.5){\textcolor{white}{\Large{P$_{drop}$}}}
\put(8,11.5){\textcolor{white}{$\longleftarrow$}}

\put(80,65){\textcolor{white}{\Large{P$_{through}$}}}
\put(87,61.3){\textcolor{white}{$\longrightarrow$}}
\end{overpic}\label{fig:SandiaSEM}}
\caption{(a) A schematic showing the chip-level design of the \ac{SiN} microdisk coupled to two waveguides. (b) \ac{SEM} image of the \ac{SiN} microdisk used in these experiments.  Details can be found in the text.}
\end{figure}
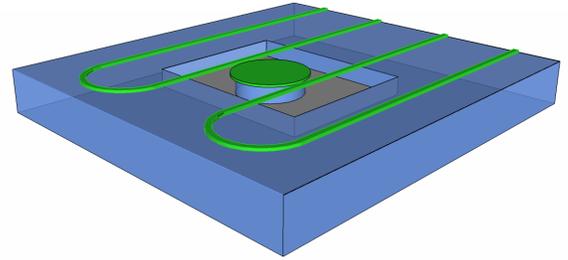
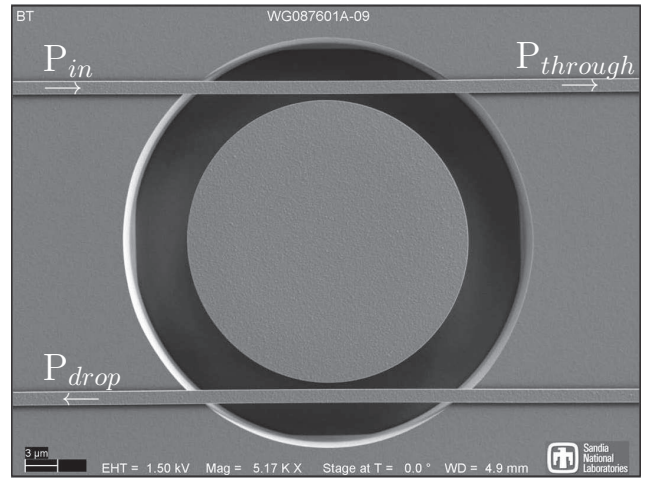

Based on these principles, groups have proposed all-optical Zeno switches employing other dissipative mechanisms, such as saturated absorption in a quantum dot coupled to a photonic crystal cavity \cite{1107.3751}, \ac{IRS} in a Silicon microdisk \cite{Wen:11B}, \ac{IRS} in an optical fiber \cite{Kieu:11} and sum and difference frequency generation in a $\chi^{(2)}$ microdisk  \cite{PhysRevA.82.063826}.  More generally, other techniques have recently been investigated to demonstrate all-optical switching with the intent of reducing operating power levels \cite{dawes2005all, bajcsy2009efficient, sridharan2011all,Salit:09, o2011all, Zhang:07}.

Here we present experimental progress towards a classical Zeno switch consisting of a \ac{SiN} microdisk embedded in hot \ac{Rb} vapor.  A key aspect of this work is the enhanced rate of \ac{TPA} that can be achieved at low power levels by confining fields to a small mode volume.  This has recently been observed at nanowatt power levels in submicron-diameter tapered optical fibers \cite{PhysRevLett.105.173602,PhysRevA.78.053803}.  Additional enhancement in the \ac{TPA} rate has been achieved by using high \ac{Q} resonators to increase the field strength and interaction time \cite{PhysRevA.80.043823,o2011all,kippenberg2004kerr}.


To demonstrate Zeno switching, we used a \ac{SiN} microdisk with a diameter of 26 $\mu$m and a thickness of 250 nm.  These devices were chosen over other high-Q microcavity designs because they have integrated waveguide coupling and a high index of refraction that supports mode compression.  The resonators contain no cladding material, allowing the evanescent field  to extend outside the device and overlap considerably with the atomic vapor.  Each microdisk was critically coupled to two waveguides in an add-drop configuration, and then coupled to four single-mode fibers using a V-groove chip with compatible spacing.  A chip-level schematic is shown in Figure \ref{fig:SandiaSchematic} and an \ac{SEM} image of the waveguide-coupled microdisk is shown in Figure \ref{fig:SandiaSEM}.

\begin{figure}
\begin{overpic}[width = \figureWidth in]%
	{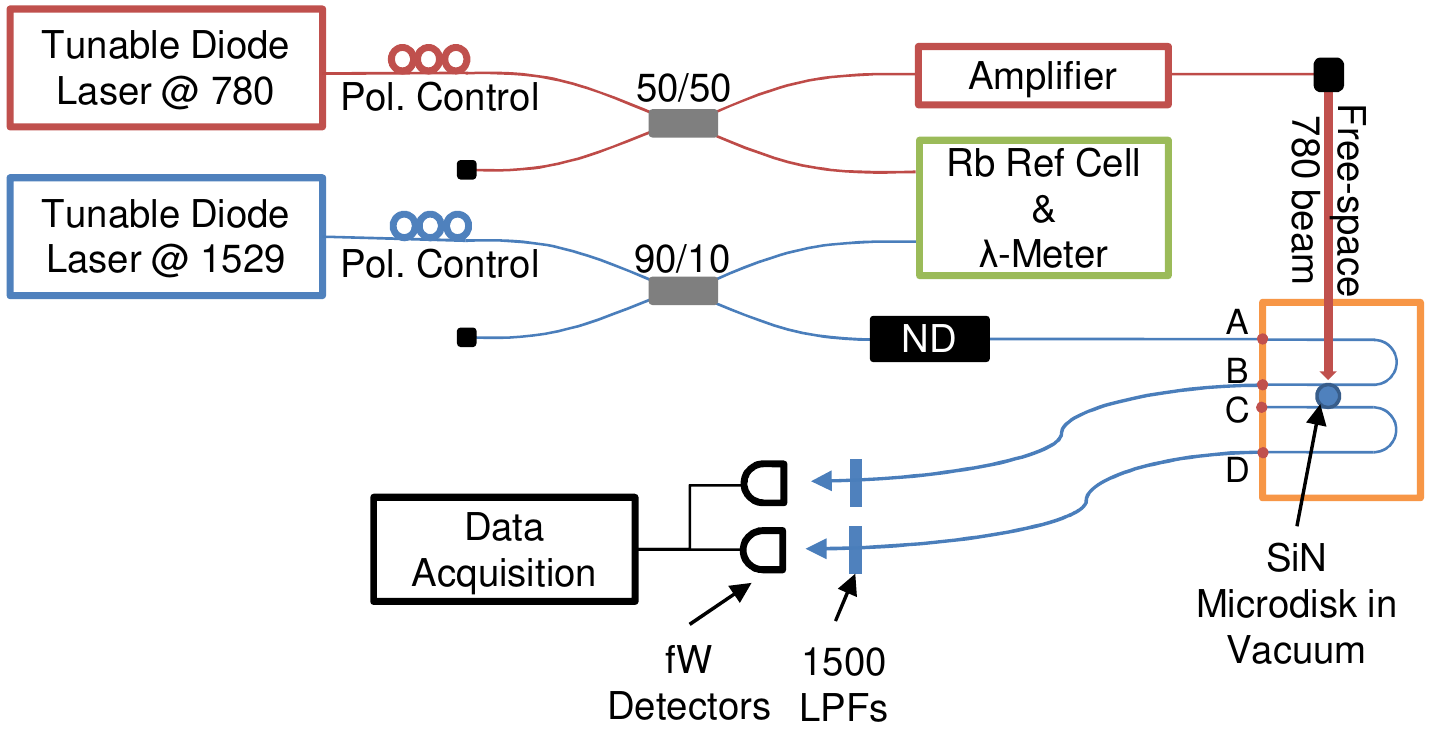}
	\put(0,0){\includegraphics[width=.8in]{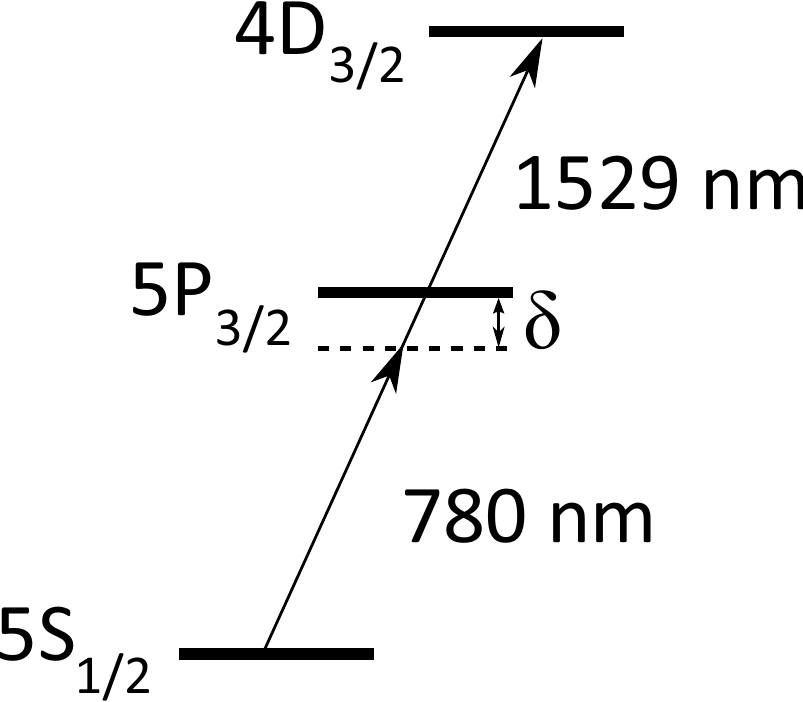}}
	\end{overpic}
\caption{\label{fig:OptDiagram} Optical diagram for the experiments described in the text.  Not shown is a 25 mW 405 nm laser focused on the resonator for thermal tuning purposes.}
\end{figure}

The setup used in this experiment is shown in Figure \ref{fig:OptDiagram}.  Two frequency-stabilized external-cavity diode lasers were used, one at 780 nm and one at 1529 nm.  Both of these beams passed through fiber splitters and into a reference cell as well as a vacuum system containing a microdisk.  The two beams were focused through a \ac{Rb} reference cell ($^{87}$Rb and $^{85}$Rb) in a counter-propagating configuration to allow monitoring of the \ac{TPA} condition while their frequencies were measured using a wavelength meter.  The 780 nm beam passed through a low-noise tapered amplifier prior to entering the vacuum system through a view-port where it was focused onto the microdisk.  The 1529 nm beam remained in fiber and was coupled to the resonator via on-chip waveguides.  The 1529 nm through- and drop-ports were measured using femtowatt detectors after passing through 1500 nm long-pass filters.

The microdisk was mounted in a vacuum system with the fibers entering through a Teflon feedthrough \cite{Abraham:98}.  The vacuum system was lightly baked to 120\C until pressures reached $10^{-8}$ Torr and then cooled to about 80\C.  \ac{Rb} vapor was injected into the system using an array of getters.  The atomic density was estimated using a fit to the tails of the absorption spectrum of a probe beam.  This method resulted in an average density between $10^{11}$ and $10^{12}$ cm$^{-3}$.  The section containing the microdisk was surrounded by view-ports to allow optical access from the top, as will be discussed below.  

Our scheme uses the two-photon transition $5S_{1/2} \rightarrow 5P_{3/2} \rightarrow 4D_{3/2}$ that consists of the well-known $D_{2}$ line near 780 nm followed by a transition near 1529 nm.  Because coupling gaps between the waveguide and the resonator were designed for wavelengths near 1529 nm, the 780 nm transition was excited using a focused free-space beam.

The matching of the \ac{cavRes} to the \ac{1529Res} was accomplished using a number of steps.  Initially, the approximate diameter of the microdisk was chosen to optimize the mode volume according to \cite{PhysRevA.79.063830} with \ac{cavRes} near \ac{1529Res}.  The diameter was then varied incrementally for a series of disks laid-out across each substrate such that a device could be selected after fabrication, coupled to fibers and secured using vacuum-compatible, high-temperature UV-curable epoxy.  Once the devices were inserted into the vacuum chamber they experienced a downward shift in \ac{cavRes} due to the change in index.  It has also been observed that \ac{SiN} devices undergo an upward shift in \ac{cavRes} when exposed to \ac{Rb} in vacuum as a result of accumulation \cite{barclay:131108} although this change can be reversed by rinsing the device in distilled water.  We have observed that this \ac{Rb} shift in \ac{cavRes} tends to increase with the maximum vapor density achieved and have used it as a tool to shift the resonance by as much as 7 nm.  The resonance can also be shifted down by etching with \ac{HF} \cite{barclay:131108}.  Ultimately, each of the aforementioned shifts were compensated for and then final tuning was accomplished thermally with a 25 mW 405 nm laser focused on the resonator.

\begin{figure}[h]

\begin{overpic}[width = \figureWidth in]{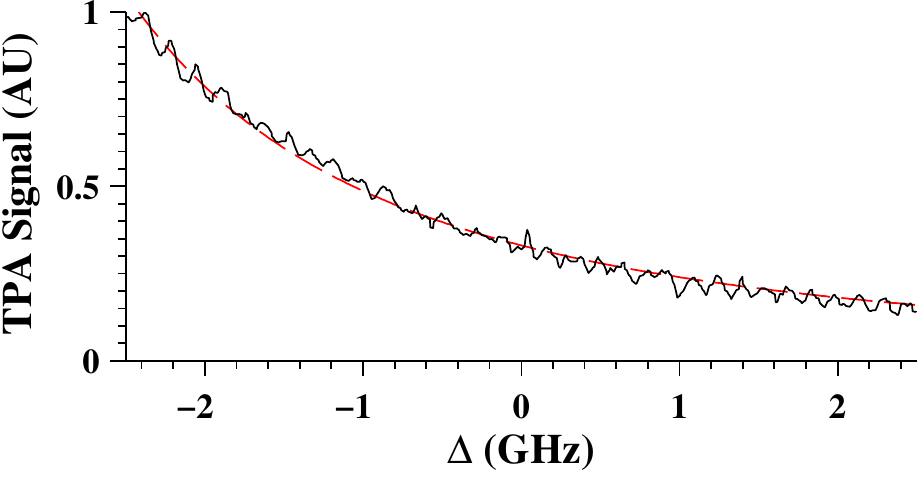}
\put(15,18){Fit: \Large{$\frac{\alpha}{(\Delta + f_0)^2}$}}
\end{overpic}

\caption{\label{fig:TPA} A typical profile of the \ac{TPA} signal in the reference cell as a function of intermediate state detuning of the 1529 nm beam.  The frequencies of both lasers were scanned simultaneously such that their total energy remained fixed with the $^{85}$Rb F=3 $5S_{1/2} \rightarrow 5P_{3/2} \rightarrow 4D_{3/2}$ transition.  The typical cavity position during data collection ($\Delta = 0$) was approximately equal to $f_0=$ 6 GHz.}

\end{figure}

While it has been predicted that classical Zeno switching could result in excellent signal contrast \cite{PhysRevA.79.063830}, the \ac{Q} of this device ($\approx 10^{5}$) is currently below what is needed for optimal performance.  With expectations of a somewhat modest signal we designed a data collection scheme to reduce noise through averaging.  A calibration algorithm was used to vary the atomic intermediate state detuning ($\delta_{Rb}$) by scanning the frequencies of the 780 nm and 1529 nm lasers by an equal and opposite amount such that their total energy remained fixed and resonant with the two-photon transition described previously.  This allowed a full scan across the profile of the cavity resonance with the conditions for \ac{TPA} satisfied, although the strength of the absorption varied as a function of intermediate state detuning.  As a control, each resonant \ac{TPA} scan was followed by a scan with the frequency of the 780 beam offset by 10 GHz such that the cavity resonance was measured in the absence of \ac{TPA}.  To compensate for any thermal drift when averaging data trials, each cavity profile was zeroed to a new frequency axis ($\Delta$) by fitting to a Lorentzian and then adding a frequency offset ($f_{0}$) to the atomic intermediate state detuning ($\delta_{Rb}$),

\begin{equation}
\Delta = \delta_{Rb}-f_{0}.
\end{equation}

Figure \ref{fig:TPA} shows the strength of the \ac{TPA} rate in the reference cell under conditions typical of each trial.  As can be seen in the plot, the \ac{TPA} condition is maintained throughout a scan of approximately 5 GHz.  The shape of the graph corresponds to the well-known  $\delta_{Rb}^{-2}$ dependence predicted by perturbation theory \cite{PhysRevA.79.063830}, as shown by the fit.

\begin{figure}[ht]
\includegraphics[width = \figureWidth in]{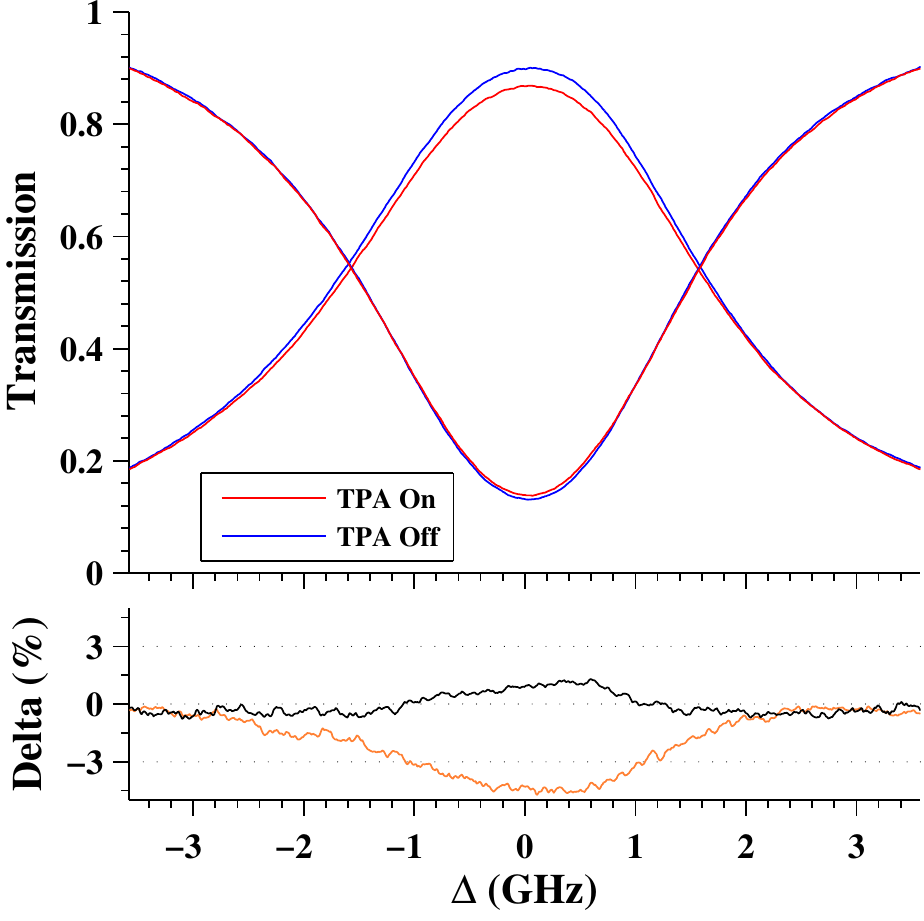}
\caption{\label{fig:Switching} Switching data: The upper plot shows cavity transmission with the conditions for \ac{TPA} satisfied in red and a control in blue.  The lower plot shows the difference in the signals.}
\end{figure}


An example of Zeno switching is shown in Figure \ref{fig:Switching}.  Approximately 100 trials were averaged for each condition.  The blue lines in the upper plot show the cavity drop-port (peak) and through-port (dip) with the 780 nm laser detuned so that the conditions for \ac{TPA} were not satisfied.  The red lines in the upper plot show the cavity response in the presence of \ac{TPA}.  The difference in these two situations is highlighted in the lower plot, where it is emphasized that the effect of the additional absorption is to increase the transmission of the through-port and decrease the transmission of the drop-port, consistent with a low-contrast Zeno switching process.

\begin{figure}[h]
\includegraphics[width = \figureWidth in]{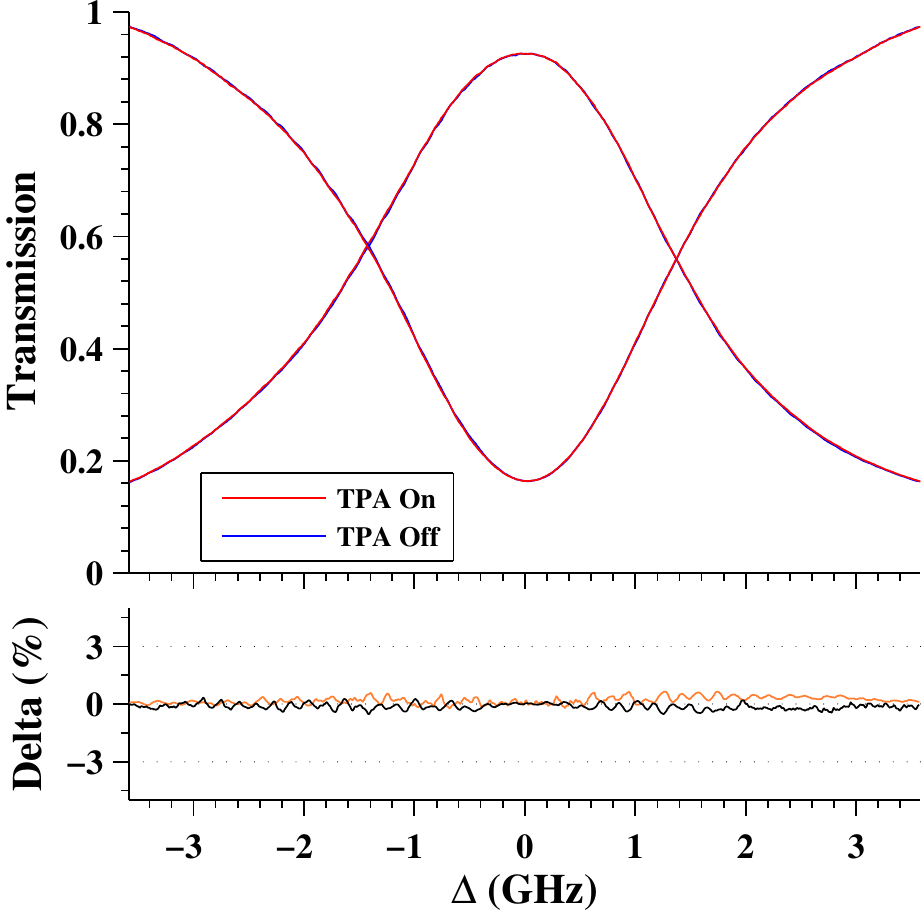}
\caption{\label{fig:Control} Control data demonstrating that no similar change is seen when the experiment is repeated with the frequency of the 1529 nm beam in a different range, as described in the text.}
\end{figure}

Figure \ref{fig:Control} shows a control case with the only difference being the absence of \ac{Rb} vapor.  As can be seen in the data, there is no similar change in the transmission of the cavity to suggest a switching event.  Although not shown here, additional controls were also performed to validate our data.  For example, in one test the cavity was thermally detuned from two-photon resonance with the power levels, density and frequency of the 780 nm free-space beam unchanged.  In another, the position of the 780 nm beam was moved from the resonator to the waveguides to determine whether absorption before or after the resonator was an issue.  All of our controls supported the conclusion that switching was due to \ac{TPA} in the evanescent field of the microdisk.  

Assuming symmetric fiber-coupling losses, the 1529 nm power in the input waveguide can be estimated to be roughly 14 nW for the switching data shown in Figure \ref{fig:Switching}.  The control data shown in Figure \ref{fig:Control} corresponds to 56 nW.  The power in the free-space 780 nm beam was 15 mW for both cases.  The intensity of the 1529 beam in the resonator was 490 W/cm$^{2}$ assuming a mode volume of 1.9 $ \times10^{-11}$ cm$^{3}$ (42 $(\lambda/n)^3$).  The intensity of the free-space beam (470 W/cm$^{2}$) would correspond to 13 nW in the input waveguide.

The primary factor that affects switching contrast is the difference between the coupled cavity Q with and without \ac{TPA}.  One way to improve the contrast of the switch is to increase the intrinsic quality factor allowing the total system Q to be increased by decreasing the waveguide coupling strengths proportionally, leaving the coupling regime unchanged.  Higher \acp{Q} in \ac{SiN} devices have been demonstrated \cite{barclay:131108} using a variety of techniques and we are optimistic that we can improve the results presented here.  We are also investigating the suitability of other types of micro-cavities for use in Zeno switching experiments, weighing factors such as intrinsic quality factor and mode volume with practical issues such as chip-scale integration and robustness.

The primary advantage of using an atomic vapor as compared to trapped atoms is the increased number of atoms in the mode and the practical advantages associated with not working at low temperature.  The disadvantages of using atomic vapors are the effects of the atomic velocities and the buildup of the atoms on the surface of the device.  The atomic velocities reduce the average time each atom spends in the mode, resulting in time-of-flight broadening \cite{bagayev1994saturated}.  In schemes such as the one demonstrated here, where the energies of the two atomic transitions are different, Doppler-free schemes are not effective and Doppler broadening dominates.  A Doppler-free scheme could be performed using the $5S_{1/2} \rightarrow 5P_{3/2} \rightarrow 5D_{5/2}$ transition in \ac{Rb}, for example \cite{PhysRevA.79.063830,olson2006two,PhysRevA.81.053825}.  The atomic build-up on the surface of the device has been used as a tool in this experiment but it also results in slight degradations in the \ac{Q} that may become more important with lower-loss resonators.  This coating on the surface of the device may also decrease the fraction of the mode that overlaps with the atomic vapor thereby reducing the effect of the atoms.  Investigations are underway to determine the extent of these effects on switching performance.


To our knowledge, these results represent the first published experimental demonstration of Zeno optical switching
based on TPA.  Using estimates based on the power of the free-space beam, all-optical switching in the low nanowatt range seems attainable.  Although the switching contrast is currently low, we believe the performance can be improved by addressing challenges related to device physics.

Funding was provided by the DARPA ZOE program (Contract No. W31P4Q-09-C-0566).


\bibliography{literatureSwitching}

\begin{thebibliography}{24}
\expandafter\ifx\csname natexlab\endcsname\relax\def\natexlab#1{#1}\fi
\expandafter\ifx\csname bibnamefont\endcsname\relax
  \def\bibnamefont#1{#1}\fi
\expandafter\ifx\csname bibfnamefont\endcsname\relax
  \def\bibfnamefont#1{#1}\fi
\expandafter\ifx\csname citenamefont\endcsname\relax
  \def\citenamefont#1{#1}\fi
\expandafter\ifx\csname url\endcsname\relax
  \def\url#1{\texttt{#1}}\fi
\expandafter\ifx\csname urlprefix\endcsname\relax\def\urlprefix{URL }\fi
\providecommand{\bibinfo}[2]{#2}
\providecommand{\eprint}[2][]{\url{#2}}

\bibitem[{\citenamefont{Misra and Sudarshan}(1977)}]{misra:756}
\bibinfo{author}{\bibfnamefont{B.}~\bibnamefont{Misra}} \bibnamefont{and}
  \bibinfo{author}{\bibfnamefont{E.~C.~G.} \bibnamefont{Sudarshan}},
  \bibinfo{journal}{Journal of Mathematical Physics}
  \textbf{\bibinfo{volume}{18}}, \bibinfo{pages}{756} (\bibinfo{year}{1977}).

\bibitem[{\citenamefont{Franson et~al.}(2004)\citenamefont{Franson, Jacobs, and
  Pittman}}]{PhysRevA.70.062302}
\bibinfo{author}{\bibfnamefont{J.~D.} \bibnamefont{Franson}},
  \bibinfo{author}{\bibfnamefont{B.~C.} \bibnamefont{Jacobs}},
  \bibnamefont{and} \bibinfo{author}{\bibfnamefont{T.~B.}
  \bibnamefont{Pittman}}, \bibinfo{journal}{Phys. Rev. A}
  \textbf{\bibinfo{volume}{70}}, \bibinfo{pages}{062302}
  (\bibinfo{year}{2004}).

\bibitem[{\citenamefont{Franson et~al.}(2007)\citenamefont{Franson, Pittman,
  and Jacobs}}]{Franson:07}
\bibinfo{author}{\bibfnamefont{J.~D.} \bibnamefont{Franson}},
  \bibinfo{author}{\bibfnamefont{T.~B.} \bibnamefont{Pittman}},
  \bibnamefont{and} \bibinfo{author}{\bibfnamefont{B.~C.}
  \bibnamefont{Jacobs}}, \bibinfo{journal}{J. Opt. Soc. Am. B}
  \textbf{\bibinfo{volume}{24}}, \bibinfo{pages}{209} (\bibinfo{year}{2007}).

\bibitem[{\citenamefont{Franson and Hendrickson}(2006)}]{PhysRevA.74.053817}
\bibinfo{author}{\bibfnamefont{J.~D.} \bibnamefont{Franson}} \bibnamefont{and}
  \bibinfo{author}{\bibfnamefont{S.~M.} \bibnamefont{Hendrickson}},
  \bibinfo{journal}{Phys. Rev. A} \textbf{\bibinfo{volume}{74}},
  \bibinfo{pages}{053817} (\bibinfo{year}{2006}).

\bibitem[{\citenamefont{Jacobs and Franson}(2009)}]{PhysRevA.79.063830}
\bibinfo{author}{\bibfnamefont{B.~C.} \bibnamefont{Jacobs}} \bibnamefont{and}
  \bibinfo{author}{\bibfnamefont{J.~D.} \bibnamefont{Franson}},
  \bibinfo{journal}{Phys. Rev. A} \textbf{\bibinfo{volume}{79}},
  \bibinfo{pages}{063830} (\bibinfo{year}{2009}).

\bibitem[{\citenamefont{Sridharan et~al.}(2011)\citenamefont{Sridharan, Bose,
  Solomon, and Waks}}]{1107.3751}
\bibinfo{author}{\bibfnamefont{D.}~\bibnamefont{Sridharan}},
  \bibinfo{author}{\bibfnamefont{R.}~\bibnamefont{Bose}},
  \bibinfo{author}{\bibfnamefont{H.~K. G.~S.} \bibnamefont{Solomon}},
  \bibnamefont{and} \bibinfo{author}{\bibfnamefont{E.}~\bibnamefont{Waks}}
  (\bibinfo{year}{2011}), \bibinfo{note}{preprint}, \eprint{arXiv:1107.3751}.

\bibitem[{\citenamefont{Wen et~al.}(2011)\citenamefont{Wen, Kuzucu, Hou,
  Lipson, and Gaeta}}]{Wen:11B}
\bibinfo{author}{\bibfnamefont{Y.~H.} \bibnamefont{Wen}},
  \bibinfo{author}{\bibfnamefont{O.}~\bibnamefont{Kuzucu}},
  \bibinfo{author}{\bibfnamefont{T.}~\bibnamefont{Hou}},
  \bibinfo{author}{\bibfnamefont{M.}~\bibnamefont{Lipson}}, \bibnamefont{and}
  \bibinfo{author}{\bibfnamefont{A.~L.} \bibnamefont{Gaeta}},
  \bibinfo{journal}{Opt. Lett.} \textbf{\bibinfo{volume}{36}},
  \bibinfo{pages}{1413} (\bibinfo{year}{2011}).

\bibitem[{\citenamefont{Kieu et~al.}(2011)\citenamefont{Kieu, Schneebeli,
  Hales, Perry, Norwood, and Peyghambarian}}]{Kieu:11}
\bibinfo{author}{\bibfnamefont{K.}~\bibnamefont{Kieu}},
  \bibinfo{author}{\bibfnamefont{L.}~\bibnamefont{Schneebeli}},
  \bibinfo{author}{\bibfnamefont{J.~M.} \bibnamefont{Hales}},
  \bibinfo{author}{\bibfnamefont{J.~W.} \bibnamefont{Perry}},
  \bibinfo{author}{\bibfnamefont{R.~A.} \bibnamefont{Norwood}},
  \bibnamefont{and}
  \bibinfo{author}{\bibfnamefont{N.}~\bibnamefont{Peyghambarian}},
  \bibinfo{journal}{Opt. Express} \textbf{\bibinfo{volume}{19}},
  \bibinfo{pages}{12532} (\bibinfo{year}{2011}).

\bibitem[{\citenamefont{Huang et~al.}(2010)\citenamefont{Huang, Altepeter, and
  Kumar}}]{PhysRevA.82.063826}
\bibinfo{author}{\bibfnamefont{Y.-P.} \bibnamefont{Huang}},
  \bibinfo{author}{\bibfnamefont{J.~B.} \bibnamefont{Altepeter}},
  \bibnamefont{and} \bibinfo{author}{\bibfnamefont{P.}~\bibnamefont{Kumar}},
  \bibinfo{journal}{Phys. Rev. A} \textbf{\bibinfo{volume}{82}},
  \bibinfo{pages}{063826} (\bibinfo{year}{2010}).

\bibitem[{\citenamefont{Dawes et~al.}(2005)\citenamefont{Dawes, Illing, Clark,
  and Gauthier}}]{dawes2005all}
\bibinfo{author}{\bibfnamefont{A.}~\bibnamefont{Dawes}},
  \bibinfo{author}{\bibfnamefont{L.}~\bibnamefont{Illing}},
  \bibinfo{author}{\bibfnamefont{S.}~\bibnamefont{Clark}}, \bibnamefont{and}
  \bibinfo{author}{\bibfnamefont{D.}~\bibnamefont{Gauthier}},
  \bibinfo{journal}{Science} \textbf{\bibinfo{volume}{308}},
  \bibinfo{pages}{672} (\bibinfo{year}{2005}).

\bibitem[{\citenamefont{Bajcsy et~al.}(2009)\citenamefont{Bajcsy, Hofferberth,
  Balic, Peyronel, Hafezi, Zibrov, Vuletic, and Lukin}}]{bajcsy2009efficient}
\bibinfo{author}{\bibfnamefont{M.}~\bibnamefont{Bajcsy}},
  \bibinfo{author}{\bibfnamefont{S.}~\bibnamefont{Hofferberth}},
  \bibinfo{author}{\bibfnamefont{V.}~\bibnamefont{Balic}},
  \bibinfo{author}{\bibfnamefont{T.}~\bibnamefont{Peyronel}},
  \bibinfo{author}{\bibfnamefont{M.}~\bibnamefont{Hafezi}},
  \bibinfo{author}{\bibfnamefont{A.}~\bibnamefont{Zibrov}},
  \bibinfo{author}{\bibfnamefont{V.}~\bibnamefont{Vuletic}}, \bibnamefont{and}
  \bibinfo{author}{\bibfnamefont{M.}~\bibnamefont{Lukin}},
  \bibinfo{journal}{Phys. Rev. Lett.} \textbf{\bibinfo{volume}{102}},
  \bibinfo{pages}{203902} (\bibinfo{year}{2009}).

\bibitem[{\citenamefont{Sridharan and Waks}(2011)}]{sridharan2011all}
\bibinfo{author}{\bibfnamefont{D.}~\bibnamefont{Sridharan}} \bibnamefont{and}
  \bibinfo{author}{\bibfnamefont{E.}~\bibnamefont{Waks}},
  \bibinfo{journal}{IEEE Journal of Quantum Electronics}
  \textbf{\bibinfo{volume}{47}}, \bibinfo{pages}{31} (\bibinfo{year}{2011}).

\bibitem[{\citenamefont{Salit et~al.}(2009)\citenamefont{Salit, Salit,
  Krishnamurthy, Wang, Kumar, and Shahriar}}]{Salit:09}
\bibinfo{author}{\bibfnamefont{K.}~\bibnamefont{Salit}},
  \bibinfo{author}{\bibfnamefont{M.}~\bibnamefont{Salit}},
  \bibinfo{author}{\bibfnamefont{S.}~\bibnamefont{Krishnamurthy}},
  \bibinfo{author}{\bibfnamefont{Y.}~\bibnamefont{Wang}},
  \bibinfo{author}{\bibfnamefont{P.}~\bibnamefont{Kumar}}, \bibnamefont{and}
  \bibinfo{author}{\bibfnamefont{S.~M.} \bibnamefont{Shahriar}}, in
  \emph{\bibinfo{booktitle}{Laser Science XXV}} (\bibinfo{publisher}{Optical
  Society of America}, \bibinfo{year}{2009}), p. \bibinfo{pages}{LSWB3}.

\bibitem[{\citenamefont{O'Shea et~al.}(2011)\citenamefont{O'Shea, Junge,
  Poellinger, Vogler, and Rauschenbeutel}}]{o2011all}
\bibinfo{author}{\bibfnamefont{D.}~\bibnamefont{O'Shea}},
  \bibinfo{author}{\bibfnamefont{C.}~\bibnamefont{Junge}},
  \bibinfo{author}{\bibfnamefont{M.}~\bibnamefont{Poellinger}},
  \bibinfo{author}{\bibfnamefont{A.}~\bibnamefont{Vogler}}, \bibnamefont{and}
  \bibinfo{author}{\bibfnamefont{A.}~\bibnamefont{Rauschenbeutel}},
  \bibinfo{journal}{preprint arXiv:1105.0330}  (\bibinfo{year}{2011}).

\bibitem[{\citenamefont{Zhang et~al.}(2007)\citenamefont{Zhang, Hernandez, and
  Zhu}}]{Zhang:07}
\bibinfo{author}{\bibfnamefont{J.}~\bibnamefont{Zhang}},
  \bibinfo{author}{\bibfnamefont{G.}~\bibnamefont{Hernandez}},
  \bibnamefont{and} \bibinfo{author}{\bibfnamefont{Y.}~\bibnamefont{Zhu}},
  \bibinfo{journal}{Opt. Lett.} \textbf{\bibinfo{volume}{32}},
  \bibinfo{pages}{1317} (\bibinfo{year}{2007}).

\bibitem[{\citenamefont{Hendrickson et~al.}(2010)\citenamefont{Hendrickson,
  Lai, Pittman, and Franson}}]{PhysRevLett.105.173602}
\bibinfo{author}{\bibfnamefont{S.~M.} \bibnamefont{Hendrickson}},
  \bibinfo{author}{\bibfnamefont{M.~M.} \bibnamefont{Lai}},
  \bibinfo{author}{\bibfnamefont{T.~B.} \bibnamefont{Pittman}},
  \bibnamefont{and} \bibinfo{author}{\bibfnamefont{J.~D.}
  \bibnamefont{Franson}}, \bibinfo{journal}{Phys. Rev. Lett.}
  \textbf{\bibinfo{volume}{105}}, \bibinfo{pages}{173602}
  (\bibinfo{year}{2010}).

\bibitem[{\citenamefont{You et~al.}(2008)\citenamefont{You, Hendrickson, and
  Franson}}]{PhysRevA.78.053803}
\bibinfo{author}{\bibfnamefont{H.}~\bibnamefont{You}},
  \bibinfo{author}{\bibfnamefont{S.~M.} \bibnamefont{Hendrickson}},
  \bibnamefont{and} \bibinfo{author}{\bibfnamefont{J.~D.}
  \bibnamefont{Franson}}, \bibinfo{journal}{Phys. Rev. A}
  \textbf{\bibinfo{volume}{78}}, \bibinfo{pages}{053803}
  (\bibinfo{year}{2008}).

\bibitem[{\citenamefont{You et~al.}(2009)\citenamefont{You, Hendrickson, and
  Franson}}]{PhysRevA.80.043823}
\bibinfo{author}{\bibfnamefont{H.}~\bibnamefont{You}},
  \bibinfo{author}{\bibfnamefont{S.~M.} \bibnamefont{Hendrickson}},
  \bibnamefont{and} \bibinfo{author}{\bibfnamefont{J.~D.}
  \bibnamefont{Franson}}, \bibinfo{journal}{Phys. Rev. A}
  \textbf{\bibinfo{volume}{80}}, \bibinfo{pages}{043823}
  (\bibinfo{year}{2009}).

\bibitem[{\citenamefont{Kippenberg et~al.}(2004)\citenamefont{Kippenberg,
  Spillane, and Vahala}}]{kippenberg2004kerr}
\bibinfo{author}{\bibfnamefont{T.}~\bibnamefont{Kippenberg}},
  \bibinfo{author}{\bibfnamefont{S.}~\bibnamefont{Spillane}}, \bibnamefont{and}
  \bibinfo{author}{\bibfnamefont{K.}~\bibnamefont{Vahala}},
  \bibinfo{journal}{Phys. Rev. Lett.} \textbf{\bibinfo{volume}{93}},
  \bibinfo{pages}{83904} (\bibinfo{year}{2004}).

\bibitem[{\citenamefont{Abraham and Cornell}(1998)}]{Abraham:98}
\bibinfo{author}{\bibfnamefont{E.~R.} \bibnamefont{Abraham}} \bibnamefont{and}
  \bibinfo{author}{\bibfnamefont{E.~A.} \bibnamefont{Cornell}},
  \bibinfo{journal}{Appl. Opt.} \textbf{\bibinfo{volume}{37}},
  \bibinfo{pages}{1762} (\bibinfo{year}{1998}).

\bibitem[{\citenamefont{Barclay et~al.}(2006)\citenamefont{Barclay, Srinivasan,
  Painter, Lev, and Mabuchi}}]{barclay:131108}
\bibinfo{author}{\bibfnamefont{P.~E.} \bibnamefont{Barclay}},
  \bibinfo{author}{\bibfnamefont{K.}~\bibnamefont{Srinivasan}},
  \bibinfo{author}{\bibfnamefont{O.}~\bibnamefont{Painter}},
  \bibinfo{author}{\bibfnamefont{B.}~\bibnamefont{Lev}}, \bibnamefont{and}
  \bibinfo{author}{\bibfnamefont{H.}~\bibnamefont{Mabuchi}},
  \bibinfo{journal}{Applied Physics Letters} \textbf{\bibinfo{volume}{89}},
  \bibinfo{eid}{131108} (\bibinfo{year}{2006}).

\bibitem[{\citenamefont{Bagayev et~al.}(1994)\citenamefont{Bagayev, Chebotayev,
  and Titov}}]{bagayev1994saturated}
\bibinfo{author}{\bibfnamefont{S.}~\bibnamefont{Bagayev}},
  \bibinfo{author}{\bibfnamefont{V.}~\bibnamefont{Chebotayev}},
  \bibnamefont{and} \bibinfo{author}{\bibfnamefont{E.}~\bibnamefont{Titov}},
  \bibinfo{journal}{Laser Phys} \textbf{\bibinfo{volume}{4}},
  \bibinfo{pages}{224} (\bibinfo{year}{1994}).

\bibitem[{\citenamefont{Olson et~al.}(2006)\citenamefont{Olson, Carlson, and
  Mayer}}]{olson2006two}
\bibinfo{author}{\bibfnamefont{A.}~\bibnamefont{Olson}},
  \bibinfo{author}{\bibfnamefont{E.}~\bibnamefont{Carlson}}, \bibnamefont{and}
  \bibinfo{author}{\bibfnamefont{S.}~\bibnamefont{Mayer}},
  \bibinfo{journal}{Am. J. Phys.} \textbf{\bibinfo{volume}{74}},
  \bibinfo{pages}{218} (\bibinfo{year}{2006}).

\bibitem[{\citenamefont{Slepkov et~al.}(2010)\citenamefont{Slepkov, Bhagwat,
  Venkataraman, Londero, and Gaeta}}]{PhysRevA.81.053825}
\bibinfo{author}{\bibfnamefont{A.~D.} \bibnamefont{Slepkov}},
  \bibinfo{author}{\bibfnamefont{A.~R.} \bibnamefont{Bhagwat}},
  \bibinfo{author}{\bibfnamefont{V.}~\bibnamefont{Venkataraman}},
  \bibinfo{author}{\bibfnamefont{P.}~\bibnamefont{Londero}}, \bibnamefont{and}
  \bibinfo{author}{\bibfnamefont{A.~L.} \bibnamefont{Gaeta}},
  \bibinfo{journal}{Phys. Rev. A} \textbf{\bibinfo{volume}{81}},
  \bibinfo{pages}{053825} (\bibinfo{year}{2010}).

\end{thebibliography}

\end{document}